\documentclass[a4paper,11pt]{article}
\usepackage{pos}
\usepackage{euclid}
\usepackage{natbib}
\usepackage{amsmath}

\title{Pre-launch optical verification of the Euclid NISP instrument and comparison with simulated images}

\author*{Louis Gabarra\textsuperscript{\normalfont\textit{a,b}}, on behalf of the Euclid Consortium}

\affiliation[a]{ Dipartimento di Fisica e Astronomia ”G. Galilei”,\\ Università di Padova, 35131, Padua, Italy}
\affiliation[b]{INFN, Istituto Nazionale di Fisica Nucleare,\\ Sezione di Padova, 35131 Padua, Italy}

\emailAdd{louis.gabarra@pd.infn.it}

\abstract{To characterise the NISP (Near-Infrared Spectrometer and Photometer) instrument optical capability before the launch of the \Euclid telescope to orbit, foreseen in 2023, data analysis of ground-based tests and Monte Carlo simulations that mimic the expected NISP performance were carried out.
Pre-launch test data were analysed to assess the fulfilment of the mission specifications in terms of Point Spread Function (PSF), set at EE50(PSF)$\,\leq$\,\ang{;;0.3}, and with a spectral resolution below 16\,\AA\,\text{px\textsuperscript{$-1$}}. We also provide a first comparison between real images from the ground-based tests with simulated ones.
We confirm the high optical quality of the NISP instrument, fulfilling the mission specifications in terms of PSF and spectral dispersion with a good agreement between the different test campaigns. We validated the PSF and spectral dispersion provided by the NISP simulator, a crucial aspect to validate the consistency between real and simulated images.}

\FullConference{%
  41st International Conference on High Energy physics - ICHEP2022\\
  6-13 July, 2022\\
  Bologna, Italy
}

\begin{document}
\maketitle

\section{Context}
The \textit{Euclid} mission aims at studying the dark Universe, i.e., the dark energy and the dark matter. For this purpose, \textit{Euclid} has been designed for two cosmological probes, the weak lensing to study the cosmic shear and the galaxy clustering (GC) to constrain the equation of state of the dark energy. The GC requires an accurate redshift of $\sigma_z / (1+z) \leq 0.001$ that will be provided by the spectroscopic channel of the NISP (Near-Infrared Spectrometer and Photometer) instrument and its three grisms, one blue grism (920--1300\,nm) and two red grisms (1254--1850\,nm). In this paper, we focus on the NISP optical assessment referring to two different ground-test campaigns \citep{Waczynski2016, Barbier2018}. One held at the Laboratoire d'Astrophysique de Marseille (LAM) in January 2020, aiming at validating the performance of the NISP flight model, and the second held at the Centre Spatial de Liege (CSL) in 2021 to verify the performance of the payload module (PLM) which comprises the \Euclid instruments coupled to the \Euclid telescope and its mechanical structure. These two ground-test campaigns were performed in cold thermal vacuum conditions with the focal plane at $\sim$90\,K \citep[][]{2019costille}. A previous ground-test campaign at warm demonstrated the diffraction-limited performance of the NISP optical
assembly \citep{grupp2019}. 
We present in the following some results from the cold thermal vacuum tests made at LAM and CSL, starting from best focus determination up to the spectral dispersion of the NISP red grism, verifying the consistency between real and simulated images.

\section{Best focus determination}
\begin{figure}[ht]
 \centering
    \includegraphics[width=\textwidth]{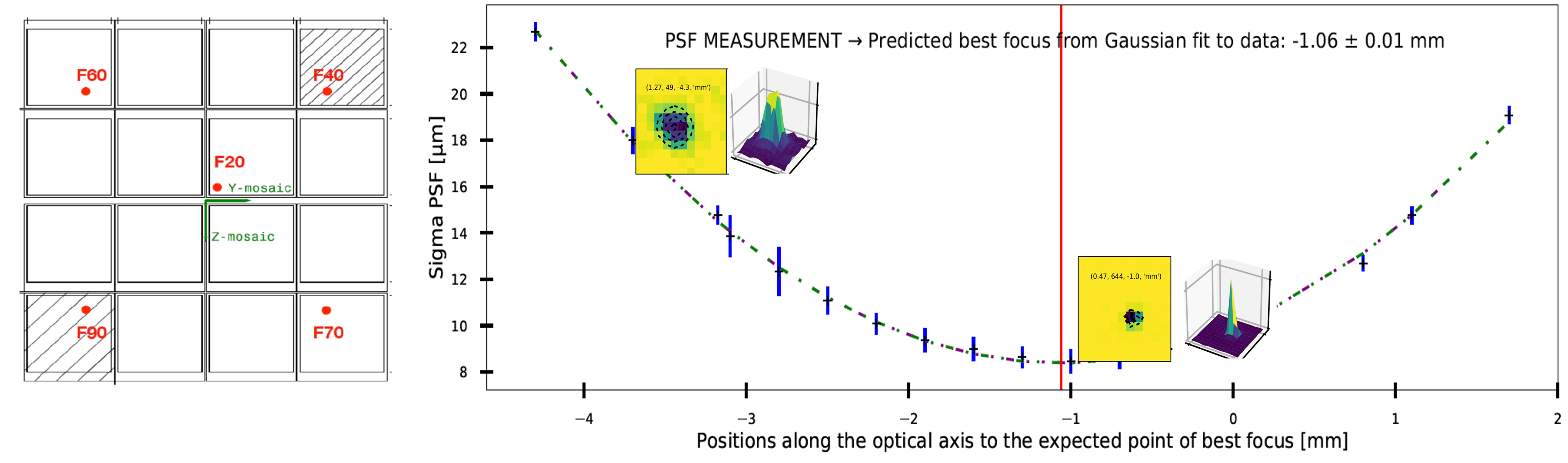}
   \caption{\emph{Left}: The different reference positions on the NISP focal plane used for PSF and spectral calibration tests. \emph{Right}: The radius containing 68\% of the flux of the PSF as a function of the NISP position along the optical axis. The reference position, i.e., zero mm, is the expected best focus position obtained from Zemax ray tracing simulation of an as-built NISP instrument model. This test was performed with a 1000\,nm point-like source using the \Euclid $Y_{\rm E}$ filter and illuminating the reference position F20.}   \label{bestfocus}

\end{figure}

A best focus determination test was performed for NISP during the ground-test campaign at LAM in 2020. To determine the best NISP focal plane position along the optical axis, the PSFs have been measured illuminating one detector with a monochromatic point-like source (see Fig. \ref{bestfocus}). 

\section{Probing the PSF on the NISP focal plane}
\begin{figure}
  \begin{minipage}[c]{0.25\textwidth}
    \centering
\includegraphics[width=\textwidth]{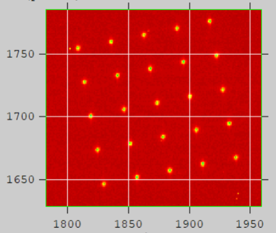}
  \end{minipage}\hfill
  \begin{minipage}[c]{0.75\textwidth}
  \centering
    \includegraphics[width=0.9\textwidth]{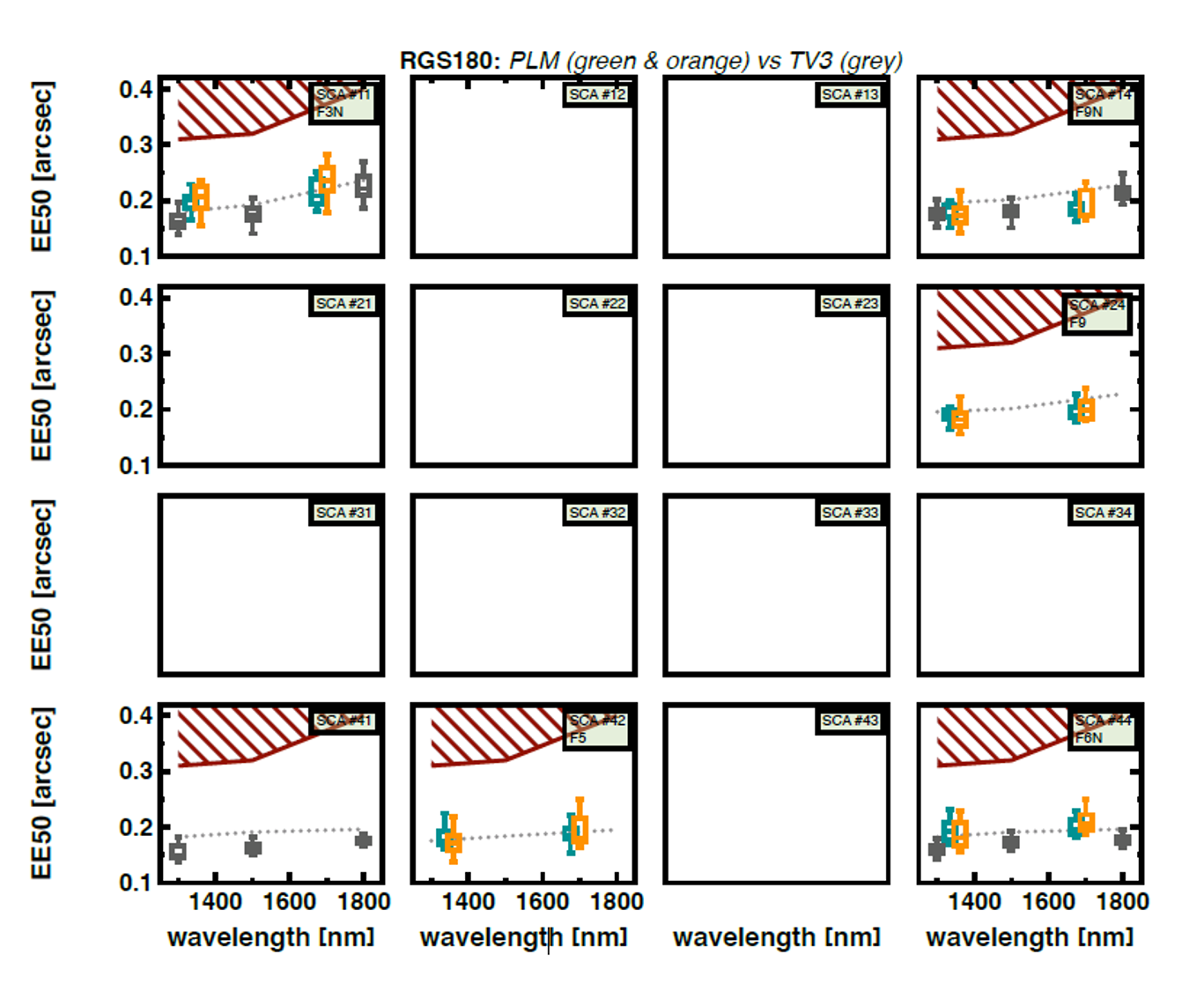}
  \end{minipage}
  \caption{\emph{Left}: PSF pattern obtained from the illumination of the focal plane with 25 sources. \emph{Right}: Results of the test at CSL using point-like sources at two different wavelengths, 1310 and 1650 nm, and for two NISP grism orientations are indicated with the blue (180$^{\circ}$) and orange (184$^{\circ}$) squares. Results of the test at LAM using 3 different wavelengths, 1300, 1500, and 1800 nm, are indicated with the grey boxes. All parameter space underneath the hatched area is consistent with performance requirements. Credit: W. Gillard.} \label{PSF_focal_plane}
\end{figure}

The PSF has been evaluated at LAM in 2020 by illuminating NISP with a monochromatic point-like source using the three \Euclid filters, i.e., $Y_{\rm E}$\,(950--1212$\,{\rm nm}$), $J_{\rm E}$\,(1168--1567$\,{\rm nm}$), and $H_{\rm E}$\,(1522--2021$\,{\rm nm}$), and the three \Euclid grisms. The two red grisms, with orientations 0$^{\circ}$ and 180$^{\circ}$, have also been tested in their additional orientations obtained by rotating the filter wheel by $-4^{\circ}$\ and\ $+4^{\circ}$\ respectively. In each configuration, the PSF was measured in at least two different wavelengths and in four positions of the focal plane, i.e., positions F40, F60, F70, F90 indicated on the left panel of Fig.\,\ref{bestfocus}.
During the ground-test campaign at CSL in 2021, only seven out of the 16 detectors were operational (see \citep{Maciaszek_2022} for the related operational issue) and the comparison between the results of the two ground-test campaigns have been done on a subset of detectors. 
We present in Fig.\,\ref{PSF_focal_plane} the results from the two campaigns for the red grism with orientation 180$^{\circ}$ and in its position obtained by rotating the filter wheel by $+4^{\circ}$. Beyond the agreement with the requirement, we can see in the figure the consistency between the results of the two ground-test campaigns. 

\section{Spectral dispersion from ground-test campaigns and from simulations}
\begin{figure}[ht]
 \centering
      \includegraphics[width=\textwidth]{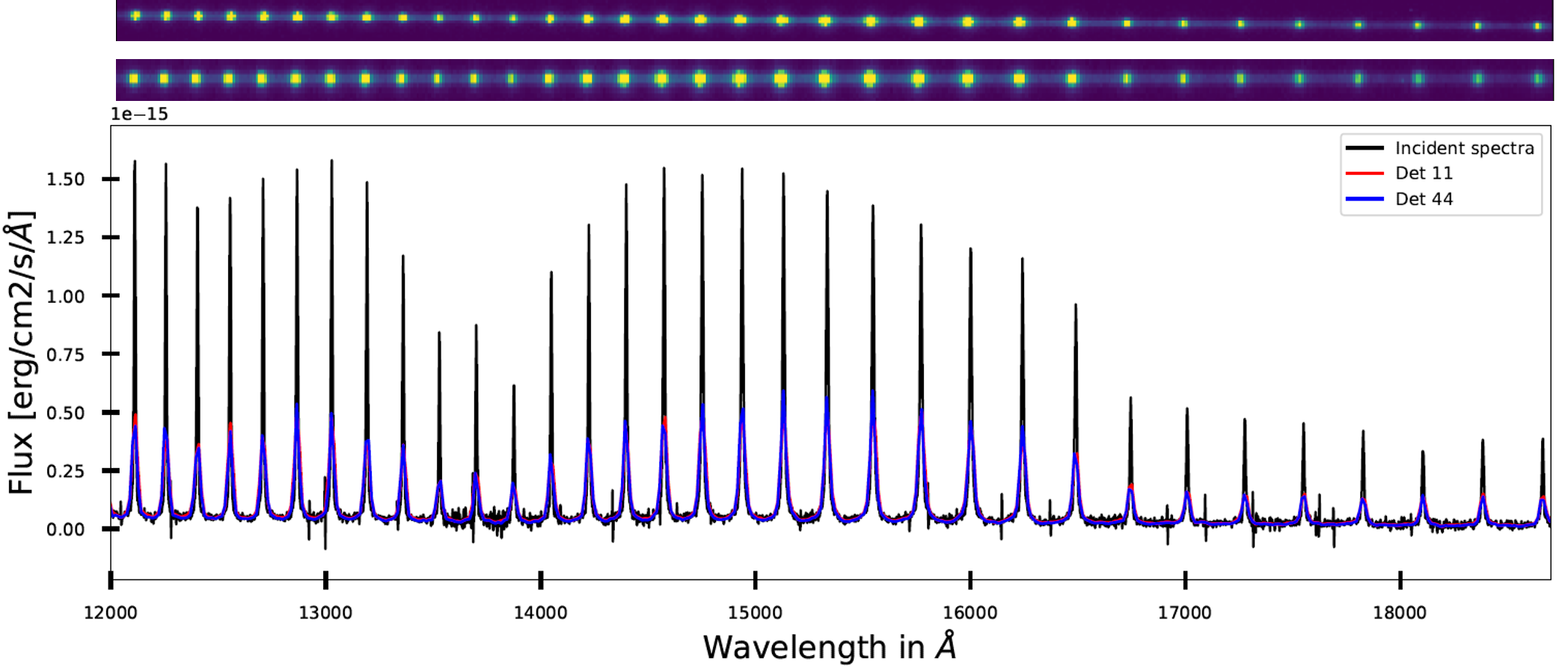}
    \caption{\emph{Top}: Ground-test image of the etalon. \emph{Middle}: Simulated image of the etalon.  \emph{Bottom}: Extracted 1D spectra from the simulated image of the etalon (coloured lines) and the true etalon spectrum (black line).} \label{dispersionEXTR}
\end{figure}

The spectral dispersion of the NISP instrument has been measured using a point-like polychromatic Fabry–Pérot etalon source. The etalon provides 33 transmission peaks in the 1254--1850 nm band pass, i.e.,\ the red grism passband. The red grism spectral dispersion has been measured on average at 13.51\,$\pm$\,0.06\,\AA\,px\textsuperscript{$-1$} (Gillard et al., in prep.), which is in agreement with the requirement set below\ 16\,\AA\,px\textsuperscript{$-1$}.\ We processed the etalon spectrum with the \Euclid pixel simulator of the NISP spectroscopic channel that we have parameterised to reproduce the conditions in laboratory, i.e., without astrophysical sources of noise. In Fig.\ \ref{dispersionEXTR}, we present the extracted 1D spectra of the simulated images to provide a preliminary validation test of the \Euclid official extraction pipeline. The agreement between the real and simulated images, and the successful extraction of the 1D spectra demonstrate that the pipeline successfully recovers the spectral dispersion.

\section{Results and conclusion}
We found a satisfying agreement between the two ground-test campaigns at LAM and at CSL, confirming that NISP exceeds PSF and spectral dispersion requirements.\ We performed a first comparison between simulated and real ground-test NISP images and found a very good agreement for both PSF and spectral dispersion measurements, attesting for a successful recovery of the NISP optical performance by the pixel simulator of the \Euclid spectroscopic channel.
\\
\\
\footnotesize
\textit{Acknowledgements}. The author thanks Chiara Sirignano, Stefano Dusini, Antonino Troja, William Gillard, and the NISP team for their help in shaping this work. \AckEC

\setlength{\bibsep}{0pt plus 0.3ex}
\bibliography{bib}

\end{document}